\documentclass{ws-procs9x6-cpt25}
\usepackage{graphicx}
\usepackage{epstopdf}

\begin{document}

\newcommand{\refeq}[1]{(\ref{#1})}
\def\etal {{\it et al.}}
%any other macros go here 

\newcommand{\beq}{\begin{equation}}
\newcommand{\eeq}{\end{equation}}
\newcommand{\bal}{\begin{aligned}}
\newcommand{\eal}{\end{aligned}}

\def\al{\alpha}
\def\be{\beta}
\def\ga{\gamma}
\def\de{\delta}
\def\ep{\epsilon}
\def\ze{\zeta}
\def\et{\eta}
\def\la{\lambda}
\def\th{\theta}
\def\ka{\kappa}
\def\rh{\rho}
\def\si{\sigma}
\def\ta{\tau}
\def\ph{\phi}
\def\ch{\chi}
\def\ps{\psi}
\def\om{\omega}

\def\Ga{\Gamma}
\def\De{\Delta}
\def\Th{\Theta}
\def\La{\Lambda}
\def\Si{\Sigma}
\def\Ph{\Phi}
\def\Ps{\Psi}
\def\Om{\Omega}

\def\cG{{\cal G}}
\def\cL{{\cal L}}
\def\cM{{\cal M}}

\def\mn{{\mu\nu}}
\def\ab{{\al\be}}

\def\sb{\overline{s}}

\def\tg{{\tilde g}}
\def\tt{{\tilde t}}
\def\tR{{\tilde R}}
\def\tr{{\tilde r}}

\def\prt{\partial}
\def\pt#1{\phantom{#1}}

\title{New Directions in Gravity Searches for Spacetime-Symmetry breaking}

\author{Q.G.\ Bailey,$^1$}

\address{$^1$Department of Physics and Astronomy, Embry-Riddle Aeronautical University,\\
Prescott, Arizona 86301, USA}

\begin{abstract}
In this talk,
we review recent results in testing spacetime symmetries in gravitational physics.
Topics discussed include new signals for hypothetical Lorentz and diffeomorphism symmetry violations
in short-range gravity tests.
We review results for multipole expansions that predict extra polarizations for gravitational waves.
Both explicit and spontaneous spacetime-symmetry breaking origins are considered.
We also discuss recent numerical results for black hole solutions in a vector field model
of spontaneous symmetry breaking.
\end{abstract}

\bodymatter

\section{Introduction}
In the last decades, 
a plethora of observational and experimental results have been obtained constraining spacetime-symmetry breaking in gravitational physics.\cite{datatables,rev1,rev2,rev3}
Tests include short-range gravity,\cite{short} 
gravimetry,\cite{gravi}
lunar laser ranging,\cite{llr}
pulsars,\cite{pulsars}
and gravitational waves (GW).\cite{gwtests}
While no statistically significant signal for spacetime-symmetry breaking has been found, 
motivation for searches come from the possibility of obtaining an hint of new physics beyond General Relativity (GR) and Standard Model of particle physics.
Indeed, 
mechanisms exist that could produce broken local Lorentz symmetry, diffeomorphism symmetry, and CPT symmetry.\cite{motiv}

Many searches for spacetime-symmetry breaking use an effective field theory (EFT) framework with generic symmetry-breaking coordinate scalars added to the action of GR and the Standard Model.\cite{sme1,k04}
These terms have tensor-indexed coefficients controlling the degree of symmetry breaking.  The premise is that spacetime-symmetry breaking is described by one or more background tensor fields coupled to known matter and fields. 
In other cases, 
specific ``toy models", 
that can generate some of these background coefficients, 
are studied for hints on how symmetry-breaking mechanisms might work, including spontaneous-symmetry breaking.

In this presentation, 
we briefly describe several recent areas for exploration in gravitational physics tests of spacetime symmetries with the author's direct involvement.
The spacetime metric signature is taken as $-+++$, 
and tensor indices are Greek letters $\mu, \nu, ...$, and other conventions can be found in the works discussed below.

\section{Short-range gravity tests}

Despite many precision experiments,
the nature of gravity is unknown on length
scales less than micrometers.
Proposals for ``fifth" forces from strong gravity interactions abound.
At submicrometer scales, 
these forces could be much stronger than
the Newtonian gravitational force,
and yet be consistent with current  experimental limits.\cite{srreview}

Lorentz-symmetry breaking signals
for short-range gravity tests 
were found using the EFT framework in the gravity sector.
In Refs.\ \refcite{bkx15} and \refcite{km17},
an approximation to first order in the coefficients was used to find the modified Newtonian force.
For example, 
the modified Newtonian potential from a point mass $m$ at the origin can be written in terms
of spherical coefficients for symmetry breaking ($k_{jm}^{N(d)lab}$) as,
\beq
U = \frac {G_N m}{r} 
+ \sum_{djm} \frac {G_N m}{r^{d-3}} 
\, Y_{jm} (\th, \ph) k_{jm}^{N(d)lab},
\label{potpert}
\eeq
where the angular dependence $\th, \ph$ is in local laboratory coordinates and $d$ labels the mass dimension of the terms in the action. \cite{km16}

The lab frame coefficients pick up time dependence when related to putative Sun-centered frame coordinates,\cite{datatables} thereby serving as a signature for experiments.
The result in equation \refeq{potpert} has already been used for analysis in experiments.\cite{short}
Recent result place limits on $14$ $k_{jm}^{N(6)}$ coefficients
and $22$ $k_{jm}^{N(8)}$ coefficients at the $10^{-9} \, m^2$
and $10^{-12} \, m^4$ levels, 
respectively.

The first order approximation makes searches 
in some short-range tests challenging, 
as some tests are designed 
to probe very small length scales 
at the cost of sensitivity to forces of Newtonian strength.
In the publication Ref.\ \refcite{bjsa23}, 
the authors found exact solutions for a subset of coefficients in the EFT framework.
Some of these coefficients are inaccessible in other tests and furthermore, 
they allow for forces larger than the Newtonian force between masses on small length scales.

In the presence of certain forms of spacetime-symmetry breaking, 
the Newtonian potential between two masses depends on two length scales and the relative amplitudes are controlled by the relative sizes of the coefficients.
The result for the Newtonian potential Green function for a point mass at $\vec r^{\, \prime}$ is 
\beq
\cG_1 = \frac {1}{2 \pi R} \Bigg[ 
1 + \frac 12 a_1 e^{\pm i w_1 R}
- \frac 12  a_2 e^{\pm i w_2 R}
\Bigg],
\label{cG1}
\eeq
where $R=|\vec r - \vec r^{\,\prime}|$.
In this expression, 
the amplitudes of term depend on coefficient ratios, e.g.,
\beq
a_1=
\left( \frac { \pm (1+2\ch) }{\sqrt{1+4\ch}}
-1 \right),
\label{amps}
\eeq
where $\ch=(k_2+k_3)/k_1$ (the $k_n$'s are short for rotational scalar coefficients like $k_1=(k^{(6)})^{0i0i0j0j}/10$, etc.). 
The ``wave numbers" $w_1, w_2$ depend also on the $k_n$'s; 
leading to damping and oscillatory behavior.\cite{bjsa23}

The new result is isotropic and so does not introduce sidereal dependence but it introduces two length scales and possibly large amplitude behavior (near $\ch=-1/4$) that could be measured in sensitive tests.\cite{decca}
It remains an open problem for experimental analysis.

\section{Gravitational wave generation and polarizations} 
\label{generation}

In one limit of the EFT approach, 
the terms in the Lagrange density for the gravity sector are constructed with the weak-field limit of gravity in mind.\cite{bk06,km16}
Thus, 
with $g_\mn = \et_\mn + h_\mn$, 
the quadratic Lagrangian terms are
\beq
\cL = -\frac {1}{4\ka} \left( h^\mn G_\mn - \sb^{\mu\ka} h^{\nu \la} \cG_{\mu\nu\ka\la} +  \tfrac 14 h_\mn q^{(5)\mu\rh \al\nu \be \si \ga} \prt_\be R_{\rh\al\si\ga} + ...\right), 
\label{EFTlag}
\eeq
where $\cG$ is the double dual of the Riemann tensor and $\ka=8\pi G_N$.\cite{bkx15}
The ellipses are higher terms in the series organized by mass dimension $d$ of the Lagrangian terms.
The leading coefficients for Lorentz violation are the dimensionless $\sb_\mn$ coefficients and the coefficients $q^{(5)\mu\rh...}$, 
the latter with mass dimension $M^{-1}$.
These coefficients have been studied in solar system, 
gravitational wave tests, and others.

In a recent work, 
the author and collaborators studied the multipole expansion of the spacetime metric and curvature far away from a radiating source, 
focusing on the $\sb_\mn$ term in \refeq{EFTlag}.\cite{bgnxs24}
The basic equation solved takes the form
\beq
{\hat K}^{\mu\nu\al\be}h_{\al\be} = \ka \ta^\mn,
\label{FEh}
\eeq
where ${\hat K}$ is a second derivative operator with Riemann symmetries
built from $\et_\mn$,
$\sb_\mn$, 
that satisfies $\prt_\mu {\hat K}^{\mu\nu\al\be} =0$.
Also in this expression, 
$\ta^\mn$ is the stress-energy pseudo tensor.

Solutions to this equation have been obtained in a particular gauge: $(\et^\mn + \sb^\mn ) \prt_\mu h_{\nu\la} =
\frac 12 \prt_\la 
( \et^\mn + \sb^\mn ) h_{\mn}$.
In a leading order in $\sb_\mn$ approximation, 
the result is
\beq
\bal
h_\mn &= 
\frac {\ka} {2\pi \sqrt{-\tg }} 
\int d^3r^\prime 
\frac {1}{\tR} \Big( 
\ta_\mn - 2 \ta^\al_{\pt{\al}(\mu} \sb_{\nu )\al} 
-\frac 12 \et_\mn (\ta^\al_{\pt{\al}\al}  - \sb_\ab \ta^\ab ) \\
& \pt{space, space, space +} 
+\frac 12 \sb_\mn \ta^\al_{\pt{\al}\al}
\Big) (\tt_R, \vec r^\prime),
\label{hLO}
\eal
\eeq
where $\tt_R$ and $\tR$ are modified retarded time and distance according to $\tt_R=t-(\tR-\sb_{0i}R^i )/(1+\sb_{00})$, 
$\tR=\sqrt{(\de_{ij}-\sb_{ij})R^iR^j + O(\sb^2)}$,
and $R^i =r^i- r^{\prime i} $.
The modified metric is $\tg^\mn=\et^\mn+\sb^\mn$.\cite{bgnxs24}

In terms of physically measurable quantities, 
we focus on the spacetime curvature components relevant for GW measurements; namely the six components $R_{0i0j}$.\cite{rev1}
These curvature components are calculated in the wave zone and to lowest order in a weak-field, slow motion expansion appropriate for gravitational waves.\cite{Blanchet13} 
We obtain answers to the same order as the usual quadrupole results of GR (of order $(v/c)^4$ in $h_{ij}$ in GR).
The results are dependent on the inertia tensor $I_{jk}=\int d^3r \, \ta^{00} r^i r^j$.
In GR, for a wave traveling in the $z=x^3$ direction, 
there are two independent polarizations corresponding to $R_{0202}-R_{0101}$ (``plus") and $R_{0102}$ (``cross"), 
where $x^1$ and $x^2$ are the coordinates for the two transverse direction.

In theories beyond GR, up to four more polarizations can exist.
For the beyond-GR polarizations, 
we find a ``breathing mode", 
and two ``vector" modes given by
\beq
\bal
R_{0101}+R_{0202} &= \frac {G}{\tr} \left[ (\sb_{tr})_{\perp ij}+ \frac 12 (\sb_{tr})_{nn} (\de_{ij} - n_i n_j ) \right] 
(\overset{(4)}{I})^{ij},\\
R_{030i} &= \frac {G}{\tr} \big[ \tfrac 12 \big( (\sb_{tr})_{in}+ \sb_{0i} \big)  (\de_{jk} - n_j n_k ) (\overset{(4)}{I})^{jk}\\
& \pt{space} + \big( (\sb_{tr})_{nk_{\perp}} + \sb_{0k_\perp } \big) (\overset{(4)}{I})^{ik}  \big],
\eal
\label{pols}
\eeq
where projections of quantities along $\hat n=\vec r/r$ are denoted with the index $n$ 
and $\perp$ indicates a projection perpendicular to $\hat n$. 
%like $(V_\perp)_i = V_i - n_i V_j n^j = V_i - n_i V_n$.
The distance $\tr$ is obtained from $\tR$ with $\vec r^\prime=0$, 
$(\sb_{tr})_{ij}=\sb_{ij}-\de_{ij}\sb_{kk}/3$, 
and the $(4)$ overset indicates the fourth $\tt_r$ derivative of the inertia tensor.
The results above are evaluated at the modified retarded time $\tt_r$ (the $\vec r^\prime=0$ limit of $\tt_R$ above).

Additional polarizations can be sought in existing and future GW data.
Some recent works have achieved results with Bayesian analysis favoring tensor over scalar modes and obtaining direct numerical bounds on the scalar tensor ratio.\cite{Takeda21}
Other possibilities for extraction of polarization content include a null-stream method.\cite{Liang24}
For a more complete list of references on this topic see
Refs.\ \refcite{bna24,GWtestrev}.

The analysis above was only for the $\sb_\mn$ coefficients, 
a subset of the more complete action \refeq{EFTlag}.
A thorough program for the coefficients for any mass dimension has recently been countenanced.\cite{PNsme}
Tests with data from the space interferometer mission LISA are also planned for the future.\cite{LISA}

\section{Results for explicit symmetry breaking}
\label{explicit}

The author and collaborators have also studied the explicit breaking limit of the $s_\mn$ coefficients and other terms in the EFT action that is {\it not} in the linearized approximation, 
in Refs.\ \refcite{abn21,bna24}.
Other authors have also studied this scenario.\cite{explicit,kl21}
In Ref.\ \refcite{bna24}, 
the explicit breaking limit of the EFT considered is described by the Lagrange density,
\beq
\bal
    \cL &= \tfrac {\sqrt{-g}}{2\kappa}
    \Big[R\left(1-e_1 u+e_2 s^\alpha_{~\alpha}+e_3 t^{\alpha\beta}_{~~\alpha\beta}\right) \\
    &+R^{\mu\nu}\left(e_4s_{\mu\nu}+e_5t_{\mu~\nu\alpha}^{~\alpha}\right)
    +e_6 t_{\alpha\beta\mu\nu}R^{\alpha\beta\mu\nu}\Big]+\cL_M,
    \label{genAct}
\eal
\eeq
where the $e_n$'s are dimensionless constants; they are introduced to distinguish the contributions from independent trace terms.
The coefficients $s_\mn$ and $t_{\al\be\ga\de}$ are defined in Ref.\ \refcite{k04}.
It is assumed that any matter ($\cL_M$) is not coupled to these coefficients.

In the explicit breaking context, 
the traced Bianchi identities $\nabla_\mu G^\mn=0$ impose severe constraints on the field equations via $\nabla_\mu T^\mn=0$, since there are no dynamical equations for the coefficients.\cite{bluhm19}
It turns out that certain choices of the constants in \refeq{genAct}
can circumvent the constraints stemming from the Bianchi identities.
For example, 
if one chooses $e_4-e_6=0$ then the constraint equation ($\nabla_\mu T^\mn=0$) can be satisfied if the Ricci scalar vanishes.  
Furthermore, 
one is then left with only a scalar combination 
$\Ph=-e_1 u+e_2 s^\al_{~\al}+ e_3 t^{\alpha\beta}_{~~\alpha\beta}$.
In fact the field equations are then, 
\beq
-G^\mn (1+\Ph) - (g^\mn \nabla^2-\nabla^\mu \nabla^\nu ) \Ph + 8\pi G_N T^\mn = 0.
\label{scalar}
\eeq
The trace of this equation gives $3 \nabla^\mu \nabla_\mu \Ph = 8\pi G_N T^\al_{\pt{\al}\al}$,
which is a wave equation for an extra scalar degree of freedom, 
arising here from explicit diffeomorphism breaking.
These field equations are identical to a special limit of 
the Brans-Dicke model with $\om =0$\cite{rev1}, with this limit ruled out by 
gravitational measurements.

If one calculates the effects of this scalar mode on gravitational waves, 
as in the previous section, an extra breathing mode is obtained.
Specifically, 
in addition to the two usual plus and cross modes, 
the curvature obtains the added contribution,
\beq
R_{0j0k} \supset
\frac {G_N}{6 r} (\de_{jk}-n_j n_k)
\prt_t^4 (n_l n_m I^{lm} -I^{ll}).
\label{breathing}
\eeq
The missing small coefficient prefactor in this expression indicates no smooth limit with GR, hence a discontinuity, 
as in massive gravity.\cite{hinter}
There is also a general discussion on the absence of a smooth GR limit in Ref.\ \refcite{kl21}.

When allowing for $e_4 \neq0$, and an arbitrary background $s_\mn$, 
GW solutions for \refeq{genAct} can be found that are not immediately 
ruled out by measurements.
These results have been worked out in Ref.\ \refcite{bna24}, 
and the results {\it do} show the persistence of an unsuppressed breathing mode, 
but the results are more complicated than the scalar results above, 
having dependence on various components of the coefficients $s_\mn$. 
% Thus they are not immediately ruled out, 
% especially considered the relative lack of concrete limits on extra polarizations discussed in previous section.

\section{Bumblebee black holes}
\label{bumblebee}

An alternative approach to understand spontaneous symmetry breaking, 
is to study specific models.\cite{ks89,kl01,jm01,kp09,abk10}
Vector models of spontaneous-symmetry breaking have been studied extensively in the last decades, 
for example, see Refs.\ \refcite{vector}.
The Lagrange density for such models is
\beq
\cL = \sqrt{-g} \left[ 
\tfrac {1}{2\ka} R  
-\frac {1}{4} B^\mn B_\mn - V +...
\right] + \cL_M,
\label{bbmodel}
\eeq
where the ellipses include nonminimal couplings to curvature, current couplings, etc.

Recently, 
a black hole solution was found in Ref.\ \refcite{casana18},
that arises from a nonminimal coupling to gravity ($\cL \sim \xi B_\mu B_\nu R^\mn$).
This led to many follow-up papers, 
including ones with numerical solutions.\footnote{See, for example, 
Refs.\ \refcite{bbBH}, and a more complete list in Ref.\ \refcite{bmw25}.}
In nearly all these works, 
it is assumed that the vector $B_\mu$ is constrained to lie at the minimum of the symmetry-breaking potential $V(B^\mu B_\mu)$, thus $V'=0$, 
where the prime is the derivative with respect to the argument.

The author and collaborators studied the case where $V' \neq 0$, 
but with the assumption of vanishing nonminimal couplings.\cite{bmw25}
One motivation for this study was to explore gravity solutions when the potential $V$ takes the special form of Kummar hypergeometric functions $M(n,2,z)$ with $z=B_\mu B^\mu/\La^2$;
such potentials have been shown to yield renormalizable and stable quantum field theories
in Refs.\ \refcite{ak05,abk10}.
Included in this work is a flat spacetime study of the modified classical electrodynamics of the model.\cite{dario}
For brevity, we focus on the solutions for the spacetime metric and vector field in a spherically symmetric and static spacetime.

For the study of black holes we used Eddington-Finkelstein coordinates ($v,r,\th, \ph$) where the metric line element takes the form $ds^2=-N(r) dv^2 + 2M dvdr +r^2 d\Om^2$.
These coordinates are useful since one can define a horizon by $N(r_h)=0$ for some radial coordinate value $r_h$.
In these coordinates the bumblebee field is $B_\mu=(B_v,B_r,0,0)$.
A version of the field equations in this metric, 
where the variables have been scaled to be dimensionless, 
is convenient for numerical solutions.
These equations take the form,
\beq
\bal
f^{\prime \prime} + \frac{2}{x}f'
&= 2 {\tilde V}' \frac{M}{N} f + k {\tilde V}' \frac{M^2}{N^2}f^2 f' x,\\
M' &= \ka \frac{f^2}{N^2} M^3 {\tilde V}' x, \\
N' &= \frac {(M^2-N)}{x} -\frac {\ka}{2} (f')^2 x
-\ka M^2 {\tilde V} x,
\eal
\label{EFeqnsdim}
\eeq
where $x=r/r_h \ge 0$, the scaled vector field is $f=B_v/\La$, 
and $k=\ka \La^2$, with $\La$ being a suitable energy scale for the bumblebee field. 
The dimensionless potential terms are $\tilde V = (r_h^2/\La^2 ) V$
and $\tilde V' = r_h^2 V'$.
The other bumblebee equation for $B_r$ turns out to provide a constraint relating $B_r$ to $B_v$ for the case $V' \neq 0$, which is already accounted for in \refeq{EFeqnsdim}.

First, we examined exact solutions to these equations exist for the case $M'=0$, revealing Schwarzschild-de Sitter and Reissner-Nordstr\"om spacetimes.\cite{bmw25}
Second, 
we studied of numerical solutions
using a near horizon analytical expansion for the functions $N,M,f$.
Since the numerical code fails for $N\rightarrow 0$,
we use a series expansion for the functions $N,M,f$ to seed numerical solutions that start slightly away from $x=1$ (the horizon).

\begin{figure}[h]
  \centering
         \includegraphics[width=\textwidth]{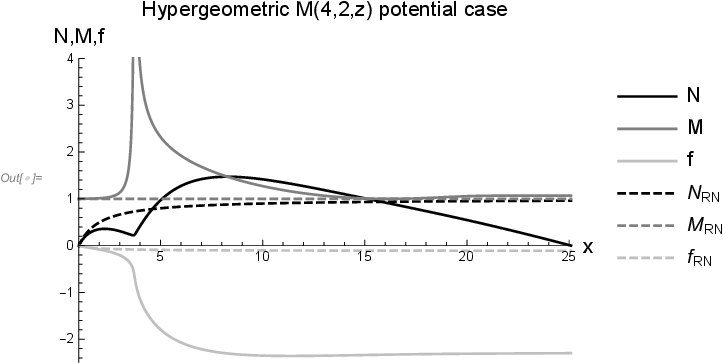}
         \caption{The hypergeometric potential ${\tilde V}={\tilde g} (M(4,2,z)+\tfrac {1}{2e^2})$ case.
         The initial horizon value used is $f'(1)=-1/9$, while ${\tilde g}=1/8$.}
         \label{horizonM4}
\end{figure}

Highlights of our findings include 
peculiar behavior of the metric functions.
We studied the quadratic potential, 
and two hypergeometric potentials.
We display a sample in Plot \ref{horizonM4}.
Also in Ref.\ \refcite{bmw25}, 
we used a series expansion method for the asymptotic region $x\rightarrow \infty$ with the variable change $u=1/x$.
These results showed that for various potential choices, naked singularity solutions, 
and solutions with a peculiar horizon-like singularities occur.
It remains an open to study these results in more detail.

\section*{Acknowledgments}
For the past works described in this proceedings, Q.G.\ Bailey was supported by National Science Foundation grants numbers 2207734 and 2308602.

\end{document}